# Efficient population transfer via bright-state

G. Grigoryan, G. Nikogosyan, T. Halfmann

*Institute for Physical Research, NAS of Armenia, Ashtarak-2, Armenia*
*Fachbereich Physik, Universitat of Kaiserslautern, Kaiserslautern, Germany*

*Propagation of short laser pulses in a medium of Λ-atoms in the intuitive sequence and under conditions of adiabatic following is studied. The regime of superluminal propagation is obtained and the conditions of efficient population transfer in the medium are analyzed.*

One of most widely used technique of population transfer in quantum systems is that of the stimulated Raman adiabatic passage (STIRAP) [1] where the counterintuitive sequence of laser pulses is realized. Recently experiments were performed where an efficient population transfer has been realized in crystals doped with rare-earth elements [2,3]. One of these experiments [3] carried out in $Pr^{3+}$:$Y_2SiO_5$ demonstrates clearly a technique termed in [3] b-STIRAP where an intuitive sequence of pulses is used. We give here a detailed theoretical investigation of this process aimed at obtaining the optimum conditions of the experiment.

As is well known, the eigenstates of the interaction Hamiltonian for a Λ-sistem, depicted in Fig.1, are:

$$|b1\rangle = \cos\varphi\sin\theta|1\rangle + \cos\varphi\cos\theta|3\rangle - \sin\varphi|2\rangle,$$
$$|b2\rangle = \sin\varphi\sin\theta|1\rangle + \sin\varphi\cos\theta|3\rangle + \cos\varphi|2\rangle, \quad (1)$$
$$|d\rangle = \cos\theta|1\rangle - \sin\theta|3\rangle.$$

where the time-dependent mixing angles and the generalized Rabi frequency are defined as $\tan\theta = \frac{\Omega_p}{\Omega_s}$, $\tan 2\varphi = \frac{\Omega}{\Delta}$, $\Omega = \sqrt{\Omega_p^2 + \Omega_s^2}$ and $\Delta = \omega_{32} - \omega_p$ is the one-photon detuning. If the fields are switched on in counterintuitive order, the bare state |1⟩ of the atoms goes adiabatically to the state |d⟩. In case of intuitive switching of fields and $\Delta \neq 0$ bare state |1⟩ adiabatically passes to the state |b1⟩ and does not leave it unless the interaction is non-adiabatic. The condition for the state |b1⟩ to be isolated from the states |d⟩ and |b2⟩ is

$$p_{ad} = \Omega^2 T / \Delta \gg 1, \quad (2)$$

where T is characteristic time of the interaction.

Propagation of the pulses is governed by Maxwell equations for slowly varying amplitudes which are written below in wave variables:

$$\frac{\partial \Omega_p}{\partial x} = iq_p a_1^* a_2, \quad \frac{\partial \Omega_c}{\partial x} = iq_s a_3^* a_2, \quad (3)$$

where $a_i$ are atomic probability amplitudes, $q_{p,s}$ -are the products of the atomic number density N and transition strengths: $q_{p,s} = 2\pi\omega_{p,s}\mu_{p,s}^2 N / \hbar c$ with $\mu_{p,s}$ being the dipole moments of the respective transitions. For simplicity we will assume in what follows that $q_p = q_s = q$. By using the atomic probability amplitudes for the state |b1⟩ and the definitions of the mixing angles, equations (3) can be rewritten as follows:

$$\frac{\partial \varphi}{\partial x} + \frac{q\cos^3 2\varphi}{4\Delta^2}\frac{\partial}{\partial t}(\sin\varphi)^2 = 0,$$
$$\frac{\partial \theta}{\partial x} - \frac{q\cos^3 2\varphi}{4\Delta^2 \sin^2\varphi}\frac{\partial \theta}{\partial t} = 0. \quad (4)$$

The obtained system of nonlinear partial differential equations may be solved analytically. The solution is

$$\varphi(x,t) = \varphi_0(\eta), \quad \theta(x,t) = \theta_0(\xi), \qquad (5)$$

where $\varphi_0(t)$ and $\theta_0(t)$ are functions given at the entrance of the medium, while $\eta$ and $\xi$ are implicit functions defined by the following expressions:

$$\eta = t - \frac{qx}{4\Delta^2}\cos^3 2\varphi_0(\eta) \qquad (6)$$

$$\int_\eta^\xi \Omega_0^2(t)dt = qx\cos^4\varphi_0(2-\cos 2\varphi_0)$$

Under the condition of large single-photon detuning, $\Delta \gg \Omega$, mixing angle $\varphi_0 \ll 1$ and the solution can be simplified:

$$\eta \approx t, \qquad \int_t^\xi \Omega_0^2(t')dt' = qx \qquad (7)$$

In this case for meeting the adiabaticity condition (2) it is necessary to require $p_{ad} = \varphi^2 \Delta T \gg 1$.

An essential distinction of the solution (7) from the regime of propagation of the counterintuitive sequence, well studied in literature [4], is that the nonlinear time $\xi$ determining the dynamics of propagation of the mixing angle $\theta$, is always larger than $t$, so we obtain superluminal propagation [5] instead of propagation at lower group velocity. Let us demonstrate this in the simplest case where $\Omega_0^2 = const$. We introduce the dimensionless optical length $z = qx/\Gamma$ with $\Gamma$ being the rate of transverse relaxation in the medium. Then it follows from expressions (5), (6) that the solution of the propagation equations (3) may be represented as

$$\Omega_p(x,t) = \Omega_{p0}(t + z\Gamma/\Omega_0^2), \quad \Omega_s(x,t) = \Omega_{s0}(t + z\Gamma/\Omega_0^2). \qquad (8)$$

Figure 2 shows the superluminal adiabaton propagation (similar ordinary adiabaton [6]) in the case where the pump pulse remains constant during the all time of interaction and is much stronger than the Stokes pulse.

The solution (5)-(6) is valid under the condition of adiabaticity not only at the medium entrance, but on the whole length of propagation. Analysis of the solution shows that the generalized condition of the adiabaticity in the medium is

$$3z\frac{\Gamma}{\Delta}p_{ad} \ll 1, \qquad (9)$$

where $p_{ad}$ is the parameter of the interaction adiabaticity for an isolated atom, defined in (2). So, the adiabaticity will be violated in the medium at rather short length, if the condition $\Gamma \ll \Delta$ is not fulfilled. The maximal length where the population transfer in the medium is possible is determined from (7) and equals $z \sim \Omega_0^2 T/\Gamma = p_{ad}\Delta/\Gamma$. At this length depletion of the pump pulse occurs and all its photons are transformed into the Stokes pulse photons. Comparing this expression with inequality (9), we see that the adiabaticity of the process is violated long before the depletion of the pump pulse.

Figures 3a and 3b demonstrate the variation of the pulse shapes and the atomic level populations in cases of counterintuitive and intuitive sequences of pulses. In both cases a considerable redistribution of photons takes place in the region of overlapping of pulses. Thereat in case of the counterintuitive sequence of pulses the leading edge of the pump pulse is gradually depleted, while at the tailing edge of the Stokes pulse an additional peak appears. In case of intuitive sequence absorption of the pump pulse occurs at the tailing edge, while the additional peak appears at the leading edge of the Stokes pulse. This dynamics of transformation of the pump photons into photons of the Stokes pulse provides reduction of the pulse propagation velocity in counterintuitive case and the superluminal propagation in intuitive case.

As follows from the analytical expressions written down and the figures shown, 100% population transfer is actually possible in both regimes. It is essential that the population transfer in intuitive sequence of pulses occurs faster and, in spite that the upper level is in the course of interaction populated really, relaxation processes have not enough time to cause significant losses.

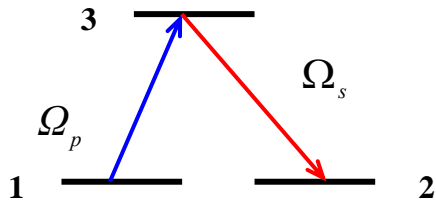

Fig1

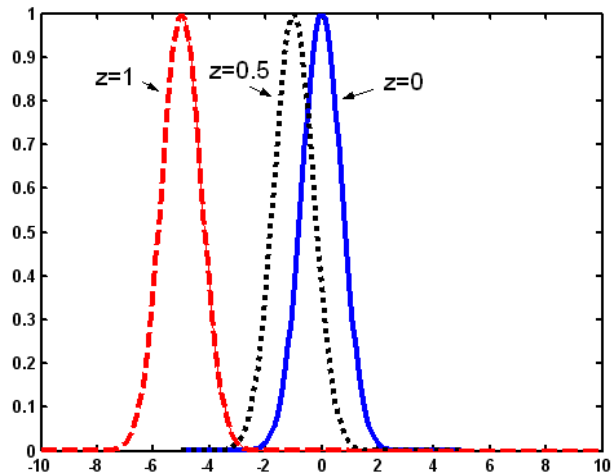

Fig.2

**Intuitive sequence**          **Counterintuitive sequence**

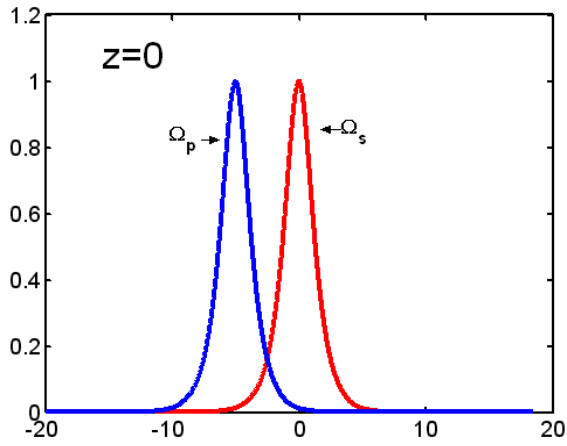
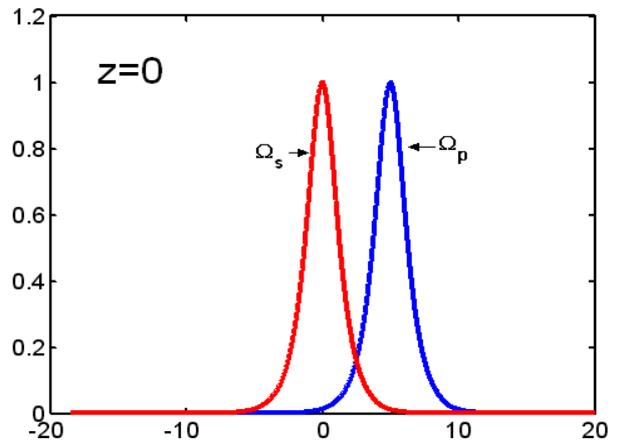
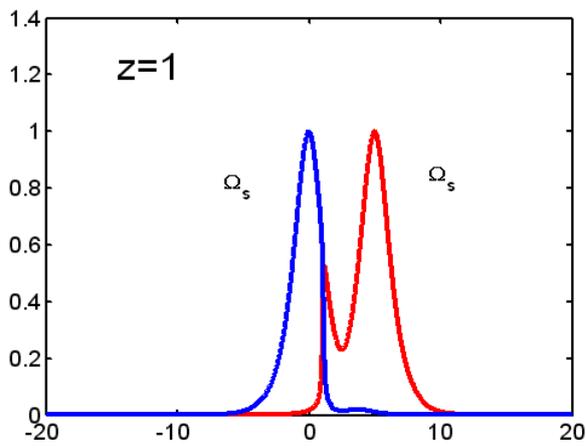
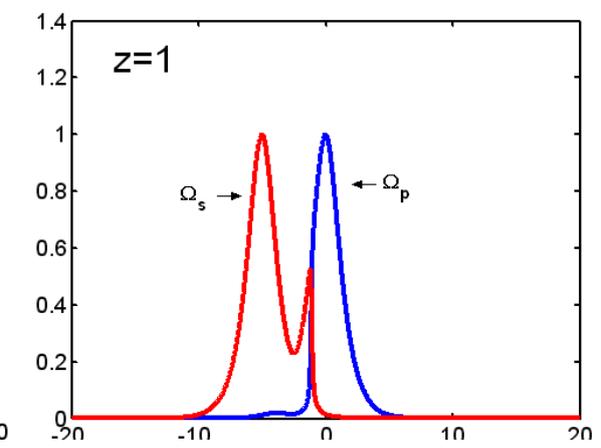
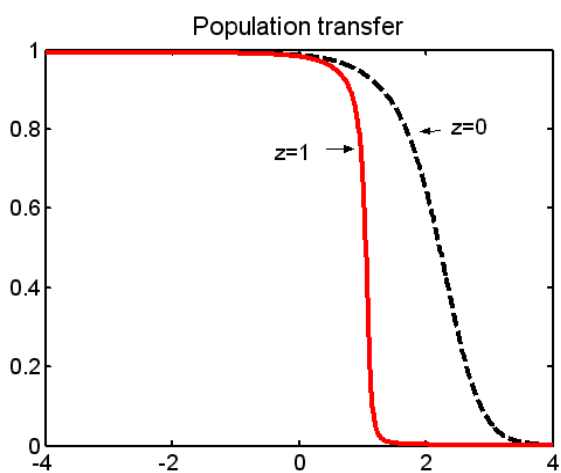
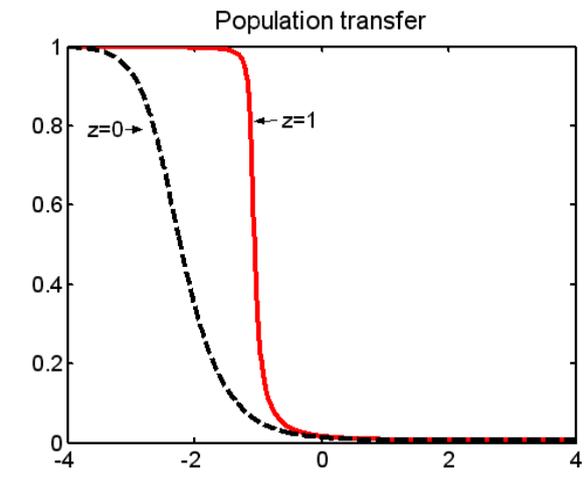

Fig3a.          Fig.3b